\documentclass[11pt,twoside]{article}
\usepackage{asp2014}

\aspSuppressVolSlug
\resetcounters

\begin{document}

\title{Exposing SED Models And Snapshots Via VO Simulation Artefacts.}

\author{Chaitra,$^1$ Sara~Bertocco,$^1$ Marco~Molinaro,$^1$ Sergio~Molinari,$^2$ Antonio~Ragagnin,$^{1,3}$ and Giuliano Taffoni$^1$}
\affil{$^1$Istituto Nazionale di Astrofisica - Osservatorio Astronomico di Trieste, Trieste, Italy}
\affil{$^2$Istituto Nazionale di Astrofisica - IAPS, Roma, Italy}
\affil{$^3$Institute for Fundamental Physics of the Universe - IFPU, Trieste, Italy}

\paperauthor{Chaitra}{chaitra.chaitra@inaf.it}{0000-0002-2484-0712}{INAF}{Via Bazzoni 2}{Trieste}{Trieste}{34124}{Italy}
\paperauthor{Sara~Bertocco}{sara.bertocco@inaf.it}{0000-0003-2386-623X}{Istituto Nazionale di Astrofisica}{Via Bazzoni 2}{Trieste}{Trieste}{34124}{Italy}
\paperauthor{Marco~Molinaro}{marco.molinaro@inaf.it}{0000-0001-5028-6041}{Istituto Nazionale di Astrofisica}{Via Bazzoni 2}{Trieste}{Trieste}{34124}{Italy}
\paperauthor{Sergio~Molinari}{sergio.molinari@inaf.it}{0000-0002-9826-7525}{Istituto Nazionale di Astrofisica}{Via Fosso del Cavaliere 100}{Roma}{Roma}{I-00133}{Italy}
\paperauthor{Antonio~Ragagnin}{antonio.ragagnin@inaf.it}{0000-0002-8106-2742}{Istituto Nazionale di Astrofisica}{Via Bazzoni 2}{Trieste}{Trieste}{34124}{Italy}
\paperauthor{Giuliano~Taffoni}{giuliano.taffoni@inaf.it}{0000-0002-4211-6816}{Istituto Nazionale di Astrofisica}{Via Bazzoni 2}{Trieste}{Trieste}{34124}{Italy}


  
\begin{abstract}
The Virtual Observatory (VO) simulation standards, Simulation Data Model (SimDM) and Simulation Data Access Layer (SimDAL), establish a framework for the discoverability and dissemination of data created in simulation projects. These standards address the complexity of having a standard access and facade for data which is expected to be multifaceted and, of a diverse range.
In this paper, we detail the realisation of an application exposing the theoretical products of one such scientific project via the simulation facades proposed by the VO. The scientific project in question, is a study of the evolution of young clusters in dense molecular clumps. The theoretical products arising from this study include a grid of 20 million SED (Spectral Energy Distribution) models for synthetic young clusters and related data products. Details on the implementation of SimDAL components in the application as well as the ways in which the data structures of SimDM are incorporated onto the existing data products are provided.
\end{abstract}

\section{Introduction}
\subsection{Simulations In The VO Context: SimDM And SimDAL}
Simulations refers to a diverse array of theoretical data. Simulations can be related to modelling astrophysical objects, simulating their evolution, interactions and usually involves codes that were used in these processes.
The efforts of VO is dedicated to regularise access of any kind of simulations and related products created in different astrophysical projects.

The main challenge in standardising access of simulations is that the data is heterogeneous in nature and the codes, polymorphic in function. VO aims to support, first, the discovery of simulations with the help of meta-data describing said simulations. The next step then, is to access the final data products with the discovered services and know-how obtained in the preliminary step.

Simulation Data Model (SimDM,  \citet{2012ivoa.spec.0503L}) describes the simulation data, providing a VO facade for theoretical data. The search and access can be facilitated with constituents of SimDAL (Simulation Data Access Layer,  \citet{2017ivoa.spec.0320L}): Repository (a registry of simulations data), Search (service to peruse or look up for a specific dataset with the proffered filter for theoretical quantities that the user is interested in) and Data Access (services to access the complete dataset or a subset). 

\subsection{The Scientific Project}
The efforts of this project have been around publishing the simulation data products of one scientific project. This scientific project is a study of the evolution of young stellar objects (YSO) clusters in massive clumps (highly dense and compact structures in clouds). A set of SED models and L/M evolutionary tracks were produced to model and then analyse the parameters associated to the compact structures\footnote{\hfill\raggedright Details of this scientific project is in the MNARS accepted paper: \url{http://vialactea.iaps.inaf.it/vialactea/eng/Deliverables/Science/5-Models/clusters_models_paper_7_mnras_accepted.pdf}.}. To be specific, the resultant data products of the study include a grid of 20 million population synthesis SED models for the protoclusters, snapshots and L/M evolutionary tracks. This data is included in the Via Lactea Knowledge Base (VLKB, a science data repository)\footnote{VIALACTEA: \url{http://vialactea.iaps.inaf.it}.}

Section 2 provides a break down on the systematisation of these data products with respect to the models proposed SimDM.  
The last section outlines some details of the application being implemented.

\section{Organising The Data Products With Respect To SimDM}
This section charts the mappings between the data products of the project to SimDM classes. Note that utypes are provided in the brackets alongside any of the first mentionings of the respective SimDM classes.
\articlefigure  {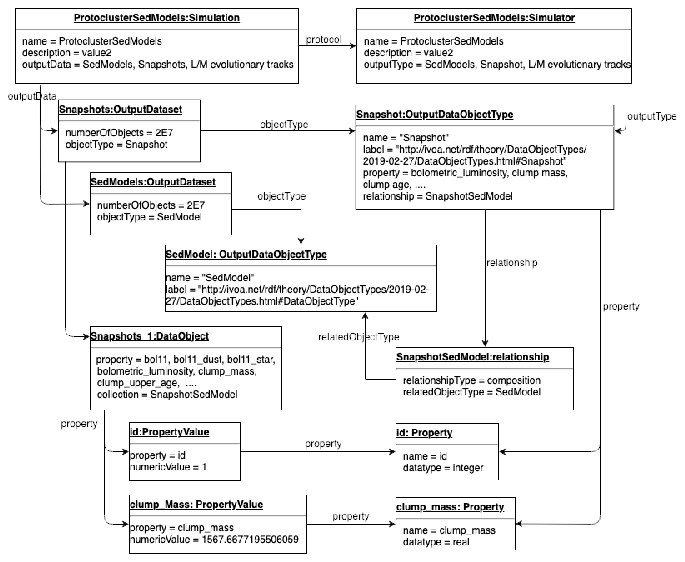}{ex_fig1}{is part of the instance diagram for the scientific project.}

Figure 1 is a part of the instance diagram for the scientific project. The experiment and protocol at the top are representative of the scientific project and the methodology used. The theoretical data products of the project are the SED models and L/M evolutionary tracks. The SED models are accompanied by snapshots which are statistical values related to the simulation data. These (including the snapshots) form the Output\-Dataset (SimDM:/resource\-/experiment\-/Output\-Dataset) of the project.

The type definitions of each individual output is given by OutputDataset.objectT\-y\-p\-e. This is a class where the OutputDataset is described along with its properties. The description of the objectType is populated in the instance of OutputDataObjectType (SimDM:/resource/pr\-otocol/OutputDataObjectType; a type of OutputDataset.objectTy\-p\-e). The outputs, for example, snapshots are represented in the instance diagram with its Output\-Data\-Object\-Type\-.label  set to ``http\-://\-ivoa\-.net\-/rdf\-/theory\-/Data\-Object\-Types\-/\-20\-19-\-02-27/\-DataObject\-Types.html\-\#Snapshot'' as currently maintained in the SKOS vocabulary. The list of properties are the parameters captured by the snapshots and the respective field contained in the model grid. An exhaustive list of these fields is omitted in the instance diagram for the sake of brevity. The snapshots are added as DataObjects (SimDM:/resource/experiment/DataO\-bject) which signify that they are direct instances of data. Each Output\-Data\-Object\-Type\-.property\- (Sim\-DM:/\-object\-/\-Property) described at the level of ObjectType is instantiated by a collection of \-Prop\-erty\-Value\- (Sim\-DM:/\-res\-ource/\-ex\-peri\-ment/\-Prop\-erty\-Val\-u\-e) objects enclosed within the snapshot`s Data\-Object instance\- (example, \-clump\_\-mass).

The snapshots represent the statistical values produced for the SED models. This relationship between the snapshots and the sed models is represented by ``SnapshotSedModel'' (a Sim\-DM:/\-object/\-Relation\-ship).

\section{Implementation Of Simulation Data Access Component}
The scope of the application lies within the context of the Data Access component of SimDAL (Repository and Search are not implemented). 
Figure 2 shows a snapshot of the application. 
\articlefigure  {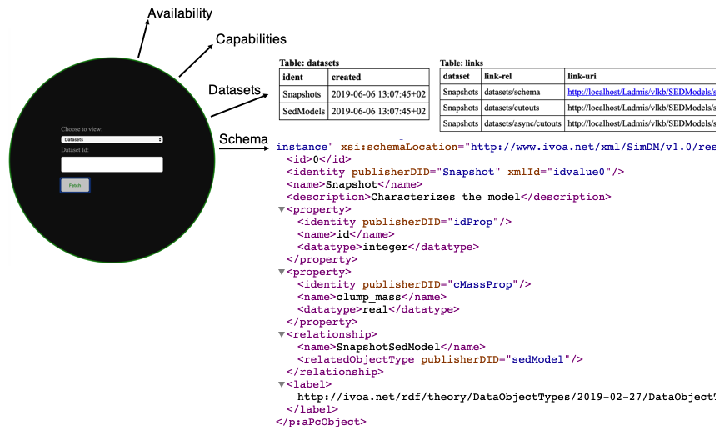}{ex_fig2}{ snapshot of the application.}

For compliancy with  SimDAL Data Access, the service must expose the datasets of the simulation alongside VOSI availability and capabilities resources (\citet{2017ivoa.spec.0524G}, VOSI chalks up the minimum requirements for a web service to be complaint to VO standards).

The application is currently in its preliminary stages of implementation and serves VOSI availability, capabilities and schemas. This is a RESTful web API and is set-up using Java technologies, SpringFramework (a Java framework) as a platform and maven (Apache Maven is a methodised building and management tool for Java based projects). 

\section{Conclusion}
In this paper, we provided a brief overview of an application being developed in compliance with the simulation artefacts of VO. The mappings between the theoretical products and the data structures proposed by SimDM are provided. 
The application is still in its preliminary stages of development.

Future work is aimed towards developing client interface and services (\{cutouts\}, \{rawdata\}, etc.) to access the statistical data as well as the SED models. We also plan to publish other simulations by extending this work; particularly into the domain of cosmological simulations.

\bibliography{P7-8}


\end{document}